\documentclass[twocolumn,aps,prl,floats,showpacs,superscriptaddress,nofootinbib]{revtex4}
\usepackage[T1]{fontenc}
\usepackage[latin9]{inputenc}
\usepackage{amsmath}
\usepackage{graphicx}
\usepackage{amssymb}
\usepackage{ifthen}

\usepackage{amssymb,amsmath}
\usepackage[usenames]{color}
\usepackage{graphicx}
\newcommand{\op}[1]{\widehat{#1}}
\newcommand{\dagop}[1]{\widehat{#1}^{\dagger}}
\newcommand{\bo}[1]{{\mathbf{#1}}}
\newcommand{\mc}[1]{{\mathcal{#1}}}

\newcommand{\wb}[1]{{\overline{#1}}}

\newcommand{\etal}{~\textsl{et al.}}

\newcommand{\pderiv}[2]{\frac{\partial#1}{\partial#2}}

\DeclareMathOperator{\tr}{Tr}

\newlength{\templength}
\newcommand{\nosh}[1]{{#1}\settowidth{\templength}{#1}\hspace*{-\templength}}

\newcommand{\REM}[1]{\ifthenelse{0=1}{#1}{}}

\begin{document}

\title{Simulation of complete many-body quantum dynamics using controlled quantum--semiclassical hybrids}

\author{P.~Deuar}

\email{piotr.deuar@lptms.u-psud.fr}
\affiliation{Laboratoire de Physique Th\'{e}orique et Mod\`{e}les Statistiques, Universit\'{e}
Paris-Sud, CNRS, 91405 Orsay, France}
\altaffiliation[Present address: ]{Institute of Physics, Polish Academy of Sciences, Al. Lotnik\'{o}w 32/46, 02-668 Warsaw, Poland}

\date{2 September 2009}
\begin{abstract}
A controlled hybridization between full quantum dynamics and semiclassical approaches 
(mean-field and truncated Wigner) is implemented for interacting many-boson systems. 
It is then demonstrated 
how simulating the resulting hybrid evolution equations allows one to obtain the full quantum dynamics for much longer times than is 
possible using an exact treatment directly. 
A collision of sodium BECs with $1.5\times10^5$ atoms is simulated, in a regime that is 
difficult to describe semiclassically. 
The uncertainty of physical quantities depends on the statistics of the full quantum prediction. 
Cutoffs are minimised to a discretization of the Hamiltonian.
The technique presented is quite general and extension to other systems is considered.
\end{abstract}

\maketitle


The calculation of the full quantum dynamics of a many-body interacting system from the microscopic description is a 
long-standing ``difficult'' problem with potential applications in many fields of physics --- 
if only one could make it numerically tractable. The difficulty is that the size of the Hilbert space grows 
exponentially with the number of particles or orbitals, while path integral Monte Carlo is foiled by
the rapid appearance of random phases. How new headway against this problem can be made will be demonstrated below.

Outside of fully integrable systems or 1D, where MPS/DMRG-based 
methods are successful, simplified descriptions are used,  
e.g. mean-field theory, Bogoliubov diagonalization, 
long-wavelength or strong interaction expansions, and Wigner-distribution based 
``c-field'' methods\cite{tWig,pgpemethod,Blakiereview}. However, some interesting problems 
fall outside the regimes of validity of these, typically where several competing effects are important or 
there is a transition between regimes that require different approximations. 
In quantum gases this occurs with rising density when interactions between the coherent component
 and incoherent particles already become of essence during the evolution, but the gas is not yet dense enough for the c-field descriptions 
to describe it with only highly occupied modes. (See \cite{Blakiereview} for a comprehensive review of 
c-field methods and their validity). This may occur e.g. in quenches of the gas\cite{quench},
colliding BECs\cite{BECcollision,MITexpt,Heexpt},
dynamics of the cooling and trapping, shock waves and the effects of obstacles\cite{obstacle-shock}
or disorder\cite{pavloff}.

This kind of dynamics is often amenable to phase-space approaches that randomly sample
the full quantum dynamics, such as positive-P\cite{pp}, stochastic wavefunctions\cite{stochwavef}, 
and stochastic gauges\cite{gauge}. They are successful when collective behaviour 
is important, but interactions between individual particles are not too strong. 
The density matrix $\op{\rho}$ of the system is re-described in terms of a probability distribution $\op{\rho} = 
\int P(\vec{v}) \op{\Lambda}(\vec{v})d\vec{v}$ of basis operators $\op{\Lambda}$
that is subsequently randomly sampled. 
These samples $\vec{v}$ are then evolved according to stochastic evolution equations that are chosen to keep the 
entire quantum dynamics of the microscopic description. A serious limitation is the ``noise catastrophe'': 
After some finite time, an exponential (or faster) growth of the noise variance occurs, 
imposing a maximum feasible simulation time $t_{\rm sim}$\cite{deuarjpa}. 
While some phenomena can be simulated\cite{deuarprl,karennacoll,otherppcalc}, 
an extension of $t_{\rm sim}$ is much sought-after, and will be demonstrated here. 


The underlying reasons why phase-space methods can overcome the Hilbert space complexity, are that quantities of 
physical interest usually involve contributions from many particles, and that limited precision is sufficient if it is well controlled.
As in Monte-Carlo methods, there is no need to follow the amplitudes of all possible configurations as 
long as one can predict physical quantities with a \emph{well-controlled uncertainty}. However --- and now we come 
to the central idea to be demonstrated here --- this can be taken further:  There is also no true need to actually follow
the troublesome exact quantum evolution equations provided that one can still predict what they would give 
\emph{with a well-controlled uncertainty}. 

How can such a roundabout prediction be achieved? If one has at one's disposal two, or more, independent approximate methods that produce 
evolution equations ``$\mc{A}$'' and ``$\mc{B}$'' without a noise catastrophe, but which bear 
sufficient resemblance to the full quantum dynamics equations ``$\mc{Q}$'', then hybrid equations can be constructed (possibly ad-hoc) 
with a continuous blending parameter $\lambda$ in a scheme resembling
\begin{equation*}
\mc{H_A} = (1-\lambda)\mc{A}+\lambda \mc{Q}\quad;\quad
\mc{H_B} = (1-\lambda)\mc{B}+\lambda \mc{Q}.
\end{equation*}
whose details will be non-universal. Here $\lambda=1$ gives full quantum dynamics, and $\lambda=0$
the original approximate methods. The hybrids 
will still contain a noise catastrophe, \emph{but at a later time} than the full quantum treatment $\mc{Q}$. Therefore, long times  $t>t_{\rm sim}^{\mc{Q}}$ 
that are not accessible by $\mc{Q}$ will be accessible by some range of $\lambda\in[0,\lambda_{\rm max}(t)\,]$.

If a physical quantity varies smoothly, preferably monotonically, as a function of $\lambda$ for hybrid $\mc{H_A}(\lambda)$, 
then an extrapolation 
can be made to $\lambda=1$, based on several calculations in the accessible range $[0,\lambda_{\rm max}(t)<1\,]$. 
One extrapolation is not yet very convincing, however, it can be checked using the other independent hybrids $\mc{H_B}(\lambda),\dots$.
When they all agree, one has an ``interpolation between extrapolations'' that is robust 
and much more reliable. Conceptually this step is similar to comparing results obtained using different 
summation techniques in diagrammatic Monte-Carlo calculations\cite{Amherst}. 

The remainder of this letter will demonstrate this procedure on a system of colliding BECs
(schematic shown in \cite{EPAPS}).
 The parameters are chosen to be close to an early experiment 
at MIT\cite{MITexpt}, but deliberately with fewer atoms, to put the system in the dilute yet Bose-stimulated regime where 
truncated Wigner and simple quasiparticle methods fail: An $N=1.5\times10^5$ atom BEC of ${}^{23}{\rm Na}$ is prepared in an 
elongated magnetic trap with frequencies $20\times 80\times80$ Hz, at a temperature 
low enough to discount the thermal component (not unusual in experiments). A brief Bragg laser pulse coherently 
imparts a velocity kick of $2v_Q={\rm 19.64 mm/s}$ to half the atoms along the long (x) condensate axis. 
The speed of the kicked atoms is supersonic (sound velocity 
in the cloud is $\le3.1$ mm/s). The trap is simultaneously turned off so that the wave-packets collide freely, 
producing a halo of scattered atom pairs moving at speeds $\approx v_Q$ relative to the overall centre of mass. 
This scattered halo exhibits a rich behaviour, which has been the repeated focus of experiments\cite{BECcollision,MITexpt,Heexpt} 
and theory\cite{BECcollthy-early,BECcollthy,plfr,Norrie,deuarprl,karennacoll,karencollcorr}. 

The high-density regime of a similar system has been treated in detail with c-field methods in \cite{Norrie}. 
Bogoliubov expansions and/or a pair-creation simplification treat the spontaneous regime, or special cases when 
BEC evolution is negligible or speed is highly supersonic\cite{BECcollthy-early,BECcollthy}
(A stochastic Bogoliubov treatment gives promising results in broader cases\cite{unpub}).  
However, major discrepancies between predictions for halo density and correlations 
arise when BEC evolution or Bose stimulation is appreciable. 
Correlations depend on the sizes of phase grains\cite{Norrie}, which develop a complicated and poorly understood 
shape\cite{plfr,karennacoll} and dynamics\cite{deuarprl,karencollcorr,Norrie} in this case. 
Parallels to unresolved questions 
in other fields of physics have been noted, such as the ``HBT puzzle'' in heavy ion collisions\cite{HBTion}. 
Trustworthy calculations that reach the end of the collision (observed in experiments\cite{Heexpt} 
but not reached by positive-P\cite{deuarprl,karennacoll}) could shed light on all these issues. 

\begin{figure}
\begin{centering}
\nosh{\includegraphics*[width=0.5\columnwidth]{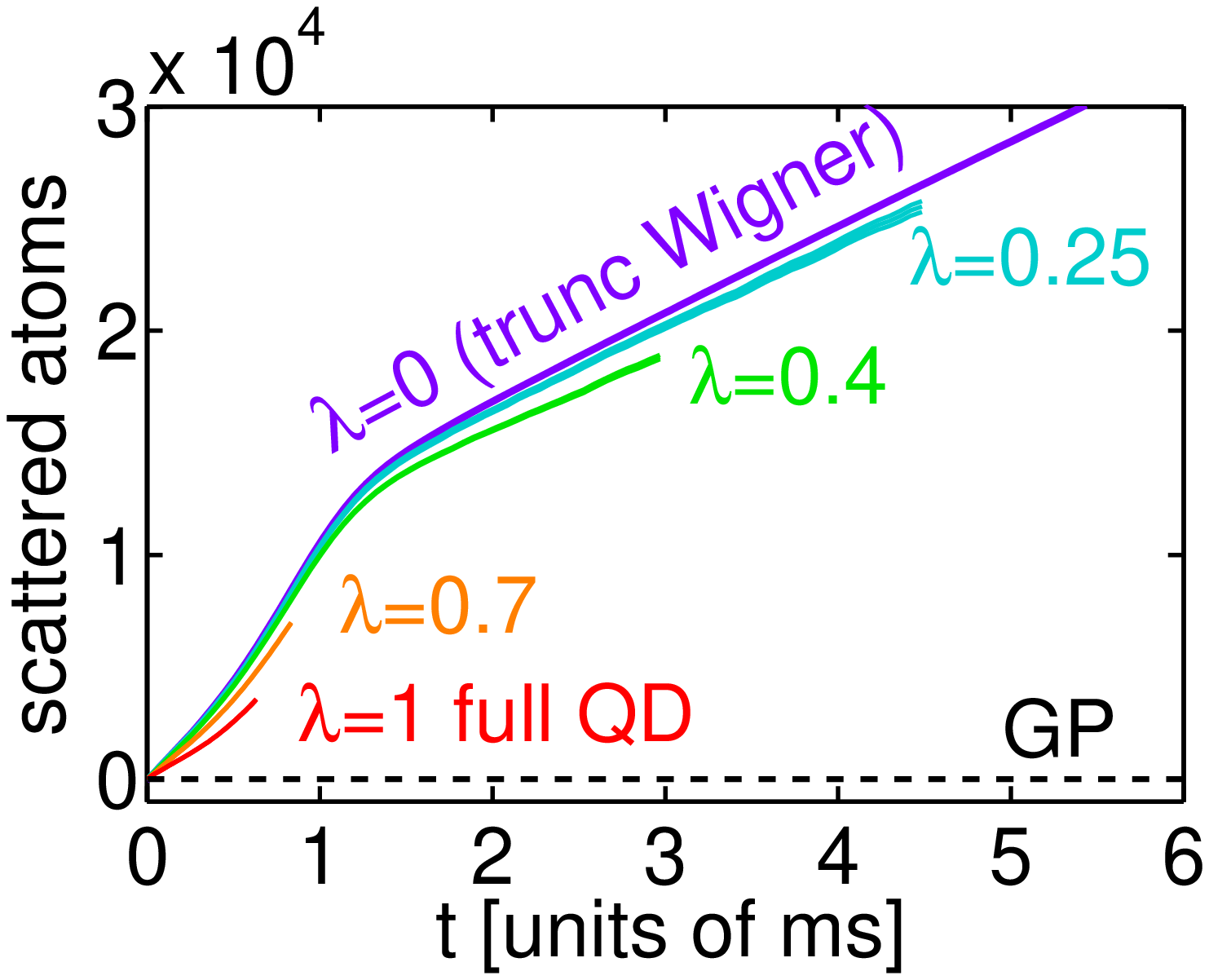}}\nosh{\raisebox{2.5cm}{\qquad\textbf{(a)}}}\hspace*{0.5\columnwidth}
\nosh{\includegraphics*[width=0.5\columnwidth]{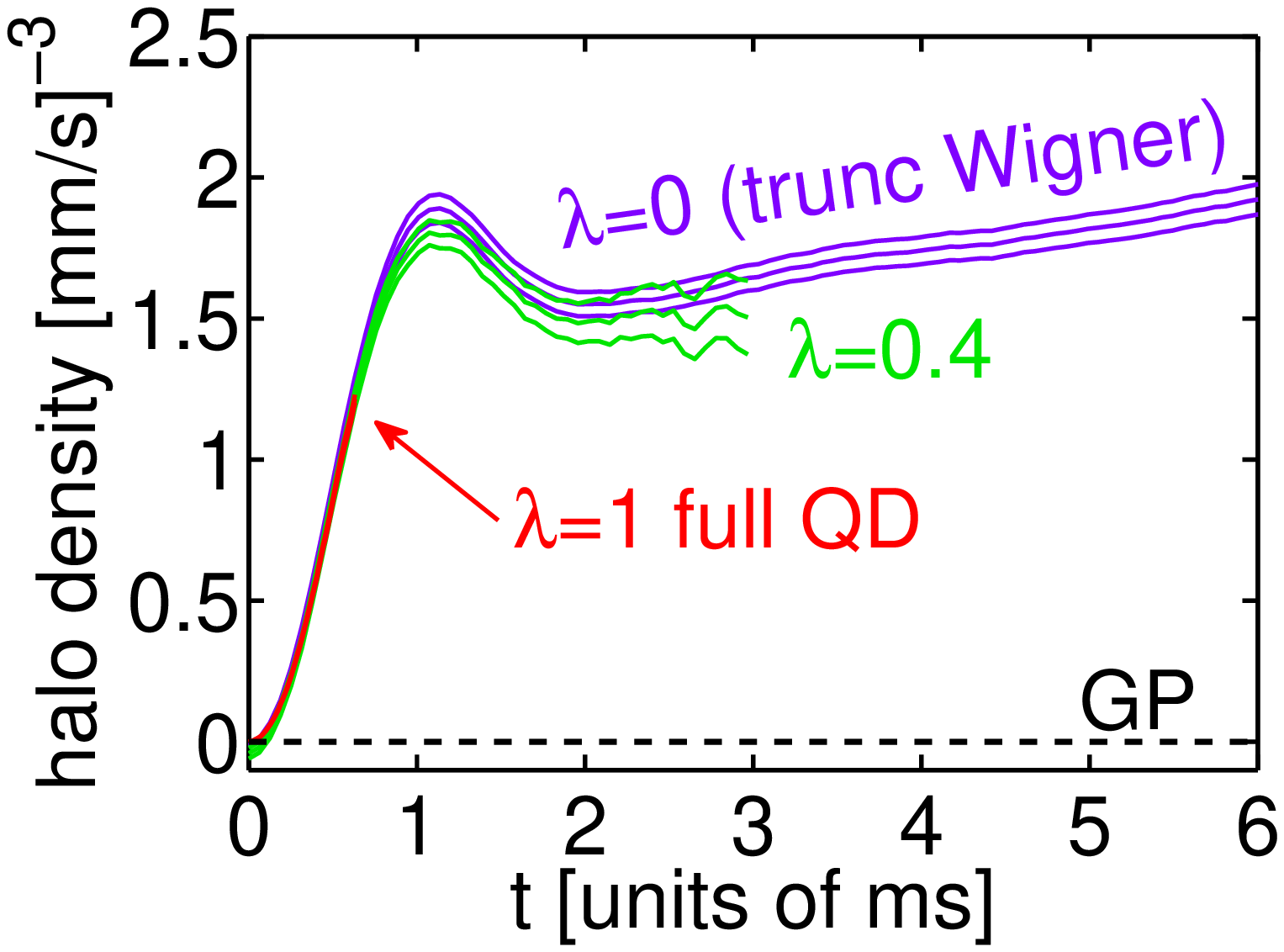}}\nosh{\raisebox{2.5cm}{\qquad\quad\textbf{(b)}}}\hspace*{0.5\columnwidth}\nosh{\mbox{}}
\end{centering}
\vspace*{-10pt}
\caption{\label{Fig-scat-nhalo} 
Wigner (purple), positive-P (red), GP (dashed) and hybrid $\mc{H_A}$ calculations at various blending parameters $\lambda$. 
(a): Total number of scattered atoms, from integration of k-space density (excluding the narrow BEC region). 
(b): Peak density of the halo (at $v_x=v_z=0$, $v_y=9.37$mm/s in velocity space).
Triple lines show 1$\sigma$ uncertainty.
}
\vspace*{-15pt}
\end{figure}

Fig.~\ref{Fig-scat-nhalo} includes predictions from  Gross-Pitaevskii (GP) mean field, truncated Wigner, and Positive-P 
calculations. The time reachable by positive-P ($t_{\rm sim}^{\mc{Q}}$) is less than a half of the collision 
time $t_{\rm coll}\approx 1400\mu$s, and both GP and Wigner give an error. The first does not treat scattering, 
while for a lattice fine enough to encompass all physics the second becomes valid only for $N\gtrsim10^6$ atoms
(one needs $\gtrsim\mc{O}(1)$ atoms per lattice site\cite{tWig}). N.b. the $k$-dependent difference between $g$ and its 
effective lattice value\cite{tWig} is $\lesssim 3\%$ here, so it has not been corrected for. 

Now let us turn to obtaining the full quantum dynamics for times longer than with the positive-P. The dynamics equations in the 
truncated Wigner, GP, and positive-P 
treatments share the GP kernel with certain additions, and turn out similar enough to 
play the role of the $\mc{A}$, $\mc{B}$, and $\mc{Q}$. 

The dynamical GP equation for the complex field $\psi(\bo{x},t)$ corresponding to the cold atom Hamiltonian 
$\op{H} = \int\!d^3\bo{x}\,\left[\dagop{\Psi}(\bo{x})H_{\rm sp}(\bo{x})\op{\Psi}(\bo{x}) + \frac{g}{2}\dagop{\Psi}(\bo{x}){}^2\op{\Psi}(\bo{x})^2\right]$ 
is 
$i\hbar\,\dot{\psi}(\bo{x}) = \left[H_{\rm sp}(\bo{x}) +g|\psi(\bo{x})|^2 \right]\psi(\bo{x})$.
An initial condensate wavefunction $\phi_{GP}(\bo{x})$ normalised to $\int d^3\bo{x}|\phi_{GP}(\bo{x})|^2  = N$ 
leads to initial conditions $\psi(\bo{x},0)=\phi_{GP}(\bo{x})$. Expectation 
values of observables $\langle\op{O}\rangle$ are calculated by making the replacements $\op{\Psi}\to\psi$ and $\dagop{\Psi}\to\psi^*$ 
in $\op{O}$. For example, the density is $\wb{n}(\bo{x}) = |\psi(\bo{x})|^2$.

In the truncated Wigner method, the dynamics is obtained by standard methods (e.g.\cite{qnoise}) based on the basis operator identities 
($\bo{x}$ dependence implied)
\begin{equation}\label{Wigop}
\op{\Psi}\op{\Lambda}=\left[\psi-\frac{1}{2}\pderiv{}{\psi^*}\right]\op{\Lambda}\ \ ;\ \ 
\dagop{\Psi}\op{\Lambda}=\left[\psi^*+\frac{1}{2}\pderiv{}{\psi}\right]\op{\Lambda}
\end{equation}
whose importance for us will be seen below. The equation of motion is as for GP but with the 
replacement $|\psi|^2\to(|\psi|^2-1)$ on the RHS. 
However, in the initial conditions the condensate field
is admixed with half a virtual particle per mode as $\psi(\bo{x},0) = \phi_{GP}(\bo{x}) + \eta(\bo{x})/\sqrt{2}$, 
where $\eta(\bo{x})$ is a 
local complex Gaussian noise with the ensemble averages 
$\langle\eta(\bo{x})\rangle = \langle\eta(\bo{x})\eta(\bo{x}')\rangle = 0$ 
and $\langle\eta(\bo{x})\eta(\bo{x}')^*\rangle = \delta^3(\bo{x}-\bo{x}')$. 
To calculate observables one ensemble averages a modified expression $f[\op{O}]$ that is obtained via
$\langle\op{O}\rangle = \tr\left[\op{O}\op{\rho}\right] = \int\!d\vec{v}P(\vec{v})\tr\left[\op{O}\op{\Lambda}\right]$
and subsequent replacements (\ref{Wigop}), which give $\int\!d\vec{v}P(\vec{v}) f(\vec{v})$. E.g. 
 $\wb{n}(\bo{x}) = \langle|\psi(\bo{x})|^2-\frac{1}{2}\rangle$.

The positive-P method uses two independent fields $\psi_1(\bo{x},t)$ and $\psi_2(\bo{x},t)$ and the identities
\begin{equation}\begin{array}{r@{ = }l@{\qquad;\qquad}r@{ = }l}
\op{\Psi}\op{\Lambda}&\psi_1\op{\Lambda} &
\dagop{\Psi}\op{\Lambda}&\left[\psi_2^*+\pderiv{}{\psi_1}\right]\op{\Lambda}, \vspace*{1pt}\\ 
\op{\Lambda}\dagop{\Psi}&\psi_2^*\op{\Lambda} &
\op{\Lambda}\op{\Psi}&\left[\psi_1+\pderiv{}{\psi_2^*}\right]\op{\Lambda}.
\end{array}
\end{equation}
The $\psi_j$ obey the Ito stochastic equations 
\begin{equation}\label{ppeq}
\hspace*{-0.3cm}\begin{array}{cccc}
i\hbar\dot{\psi}_1(\bo{x})&\!\!=&\!\!\left[ H_{\rm sp}(\bo{x}) +g\rho(\bo{x})\ -\ \,\sqrt{ig}\,\xi_1(\bo{x},t)\,\right]&\!\!\!\psi_1(\bo{x})\\
i\hbar\dot{\psi}_2(\bo{x})&\!\!=&\!\!\left[ H_{\rm sp}(\bo{x}) +g\rho(\bo{x})^*-i\sqrt{ig}\,\xi_2(\bo{x},t) \right]&\!\!\!\psi_2(\bo{x})
\end{array}\hspace*{-0.1cm}
\end{equation}
with ``complex density'' $\rho(\bo{x}) = \psi_1(\bo{x})\psi_2(\bo{x})^*$. 
Here the $\xi_j$ are delta-correlated real Gaussian noise fields with the ensemble averages 
$\langle\xi_j(\bo{x},t)\rangle=0$ and 
$\langle\xi_i(\bo{x},t)\xi_j(\bo{x}',t')\rangle = \delta_{ij}\delta(t-t')\delta^3(\bo{x}-\bo{x}')$. 
Initial conditions are $\psi_j(\bo{x},0)=\phi_{GP}(\bo{x})$ and observables are obtained with the replacements 
$\op{\Psi}\to\psi_1$ and $\dagop{\Psi}\to\psi_2^*$.

The next step will be to hybridize the truncated Wigner with the positive-P into treatment $\mc{H_A}$. It is most 
straightforward to proceed from hybrid operator identities for an off-diagonal expansion
\begin{equation}
\hspace*{-0.3cm}
\begin{array}{r@{ = }l@{\ ;\ }r@{ = }l}
\op{\Psi}\op{\Lambda}&\left[\psi_1-\frac{1-\lambda}{2}\pderiv{}{\psi_2^*}\right]\op{\Lambda}&
\dagop{\Psi}\op{\Lambda}&\left[\psi_2^*+\frac{1+\lambda}{2}\pderiv{}{\psi_1}\right]\op{\Lambda} \vspace*{1pt}\\
\op{\Lambda}\dagop{\Psi}&\left[\psi_2^* -\frac{1-\lambda}{2}\pderiv{}{\psi_1}\right]\op{\Lambda}&
\op{\Lambda}\op{\Psi}&\left[\psi_1+\frac{1+\lambda}{2}\pderiv{}{\psi_2^*}\right]\op{\Lambda}
\end{array}
\end{equation}
One obtains:
 $\wb{n}(\bo{x}) = \langle\psi_1(\bo{x})\psi_2(\bo{x})^*-\frac{1-\lambda}{2}\rangle$ and 
initial $\psi_j(\bo{x},0) = \phi_{GP}(\bo{x}) + \eta(\bo{x})\sqrt{\frac{1-\lambda}{2}}$. The usual 
truncated-Wigner-like discarding of high-order derivatives in the relevant Fokker-Planck equations, gives dynamics
\begin{equation*}
\begin{array}{cccc}
i\hbar\dot{\psi}_1(\bo{x})&\!\!=&\!\!\left[H_{\rm sp}(\bo{x})+g\rho'(\bo{x})\ -\ \,\sqrt{ig\lambda}\,\xi_1(\bo{x},t)\right]&\!\!\!\psi_1(\bo{x})\\
i\hbar\dot{\psi}_2(\bo{x})&\!\!=&\!\!\left[H_{\rm sp}(\bo{x})+g\rho'(\bo{x})^*-i\sqrt{ig\lambda}\,\xi_2(\bo{x},t)\right]&\!\!\!\psi_2(\bo{x})
\end{array}
\end{equation*}
with $\rho'(\bo{x}) = \rho(\bo{x})+\lambda-1$. 
As an aside, this corresponds to a representation based on an off-diagonal operator basis using $s$-ordered\cite{sorder} 
coherent-like states with $s=\lambda$ (See \cite{EPAPS} for details).
Fig.~\ref{Fig-scat-nhalo} shows the performance of this hybrid for several values of $\lambda$ for two halo quantities of interest. 
As desired, $\lambda<1$ calculations last for longer than the full quantum dynamics. Here the simulation time scales as
$t_{\rm sim} \approx\propto 1/\lambda$, but this is not universal. 

Hybridization of the GP and positive-P methods into treatment $\mc{H_B}$ simply entails 
replacing $\sqrt{ig}$ by $\sqrt{ig\lambda}$ in the equations (\ref{ppeq}) and following the positive-P prescription from then on. 
Here $t_{\rm sim} \propto 1/\lambda^2$.

\begin{figure}
\begin{centering}
\includegraphics[width=0.9\columnwidth]{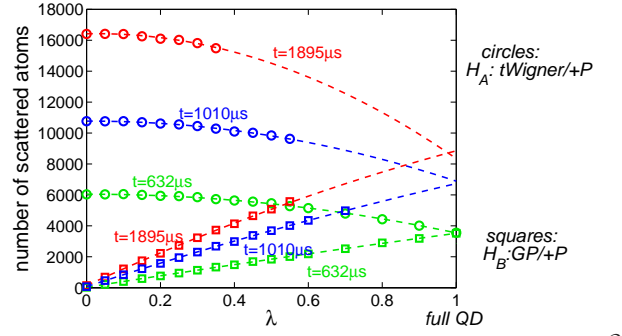}
\end{centering}
\vspace*{-10pt}
\caption{\label{Fig-extrap} 
$\lambda$-dependent predictions for several times $\ge t_{\rm sim}^{\mc{Q}}$ (symbols) and 
corresponding quadratic fits (dashed line). Fitting is via minimisation of 
rms deviation in units of $1\sigma$ data uncertainty. Data points use $\approx300-1000$ trajectories. 
}
\vspace*{-15pt}
\end{figure}

With hybrids in hand, extrapolations of the total number of 
scattered atoms to the full QD limit $\lambda=1$ 
are shown in Fig.~\ref{Fig-extrap} for several times $\ge t_{\rm sim}^{\mc{Q}}$. 
Halo peak density is in\cite{EPAPS}.

An issue here is deciding upon a fitting function -- linear, quadratic, otherwise?
Firstly, an acceptable fit must not have any statistically significant mismatch with the data.
Secondly, to exclude spurious ill-conditioned parameters, one should choose a fit that minimises 
the uncertainty in the extrapolated value at $\lambda=1$ (see below). 
One must also beware of possible stiffness in the unseen $\lambda$, and sensitivity to this is the primary 
reason why several independent hybrids are needed. Details of Fig.~\ref{Fig-extrap} are consistent with a lack of
stiffness in the unsimulated large $\lambda$ region: Firstly, for $t$ at which the whole $\lambda$ sequence is seen, there
are no inflections. Secondly, the two hybrids approach the $\lambda=1$ value from different sides but agree. 
Also, extrapolations from only a low-$\lambda$ portion of the available data should agree with ones that use the whole sequence. 
This is confirmed in  \cite{EPAPS}.

\begin{figure}
\begin{centering}
\includegraphics[width=0.8\columnwidth]{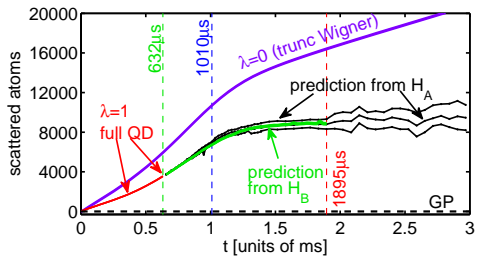}
\end{centering}
\vspace*{-10pt}
\caption{\label{Fig-predict} 
Predictions of from hybrids $\mc{H_A}$ and $\mc{H_B}$ compared with short-time full quantum dynamics 
and approximate methods. Triple lines, where visible, are $1\sigma$ uncertainty. Uses $\approx 10-20$ values of $\lambda$, as per Fig.~\ref{Fig-extrap}.
}
\vspace*{-15pt}
\end{figure}

Agreement between the $\mc{H_A}$ and $\mc{H_B}$ extrapolations in Fig.~\ref{Fig-extrap} is rather good at long times, 
but it remains to provide a well-defined uncertainty for the final prediction. Methods to obtain the statistical 
uncertainty of the $\lambda=1$ extrapolation are known\cite{numericalrecipes}. In this endeavour 
it is very helpful to know the underlying distribution of the data points $v(\lambda)$, which are ensemble 
averaged observables. Conveniently, it is known to be Gaussian by the central limit theorem, and the shown 1$\sigma$ uncertainty 
$\Delta v(\lambda)$ is its standard deviation. One rather simple way to proceed is to 
generate a number $N_S\gg1$ of ``synthetic'' data sets, where in the $j$th set 
one generates $v_j(\lambda) = v(\lambda) + \xi_j(\lambda) \Delta v(\lambda)$, with $\xi_j$ being 
Gaussian random variables of variance 1, mean zero. The synthetic data $v_j$ are distributed 
with the same mean as the original $v$ but double the variance. Now one calculates an extrapolated QD prediction $v_j(1)$ 
for $\lambda=1$ for each synthetic set $j$, and uses the distribution of these $v_j(1)$ to obtain the final uncertainty $\Delta v(1)$. 
Predictions from $\mc{H_A}$ and $\mc{H_B}$ that match within statistical uncertainty 
are trustworthy to this accuracy. The final predictions from both hybrid methods 
for the number of scattered atoms are shown in Fig.~\ref{Fig-predict}, and for halo density in \cite{EPAPS}.

One sees that the useful simulation time has been extended several-fold, allows one to reach 
the end of the collision here, and determine the total scattered atoms to be
$8800\pm400$ (at $t$=1.7ms). 
The much worse precision of the $\mc{H_A}$ result stems from the inherent vacuum noise in 
Wigner calculations and shorter segment of $\lambda$ values. However, for halo density, it is $\mc{H_B}$ that is more noisy.

Regarding limits of applicability, at very long times the uncertainty becomes 
excessive for all hybrids since the short $\lambda$ 
intervals give badly conditioned extrapolations. 
Hence the bare simulation time in the $\mc{Q}$ treatment must not be too small to ensure 
a sufficiently long $\lambda$ interval.
It is also crucial that the blending $\lambda$ enter 
the dynamics in a global way: Artificial boundaries\cite{pgpemethod,scotthoffmann} could make observables depend stiffly on the boundary position. 
For cold gases low densities can be treated perturbatively, while at high enough densities c-field treatments are valid, so that 
one expects that the blending method will be most useful at intermediate densities that ``fall through the cracks'' between these
two methods. The relative simplicity of not requiring a projection 
onto low-energy modes may also make blending appealing in other regimes.

Finally, while the emphasis has been on cold boson dynamics, the general equation-blending approach should be broadly applicable. 
For hard-core boson or fermion systems other approximations would have to be hybridised with a different complete 
phase-space description $\mc{Q}$.
One can also hybridise ``imaginary-time'' evolution for thermal equilibrium states, or 
Monte-Carlo path-integrals with the aim of predicting the ab-initio result for longer $\beta=1/T$ than is 
normally allowed by the fermion sign problem.

\emph{Concluding,} it has been demonstrated how the full quantum dynamics of a macroscopic interacting 3D system can be 
calculated for much longer times than was possible with the previously most effective method, the positive-P representation. Quantitative 
predictions for BEC collisions in the dilute stimulated regime were obtained. 
The hybrid dynamical equations used, while not actually simulating complete quantum dynamics \textit{per se}, 
can be used to confidently predict the full quantum dynamics (within a given accuracy) when several 
families of hybrids are available. 

\begin{acknowledgments}
I am grateful to Scott Hoffmann, Peter Drummond, Georgy Shlyapnikov, Boris Svistunov, Joel Corney, 
Anatoli Polkovnikov, and Evgeny Burovskiy for stimulating discussions.
This research was supported by the European Community under the contract MEIF-CT-2006-041390. 
LPTMS is a mixed research unit No. 8626 of CNRS and Universit\'{e} Paris-Sud.
\end{acknowledgments}


\clearpage

\newpage
\section*{\Large\bf Supplementary material}

\subsection{The BEC collision}

\begin{minipage}{\columnwidth}
\nosh{\includegraphics*[width=\columnwidth]{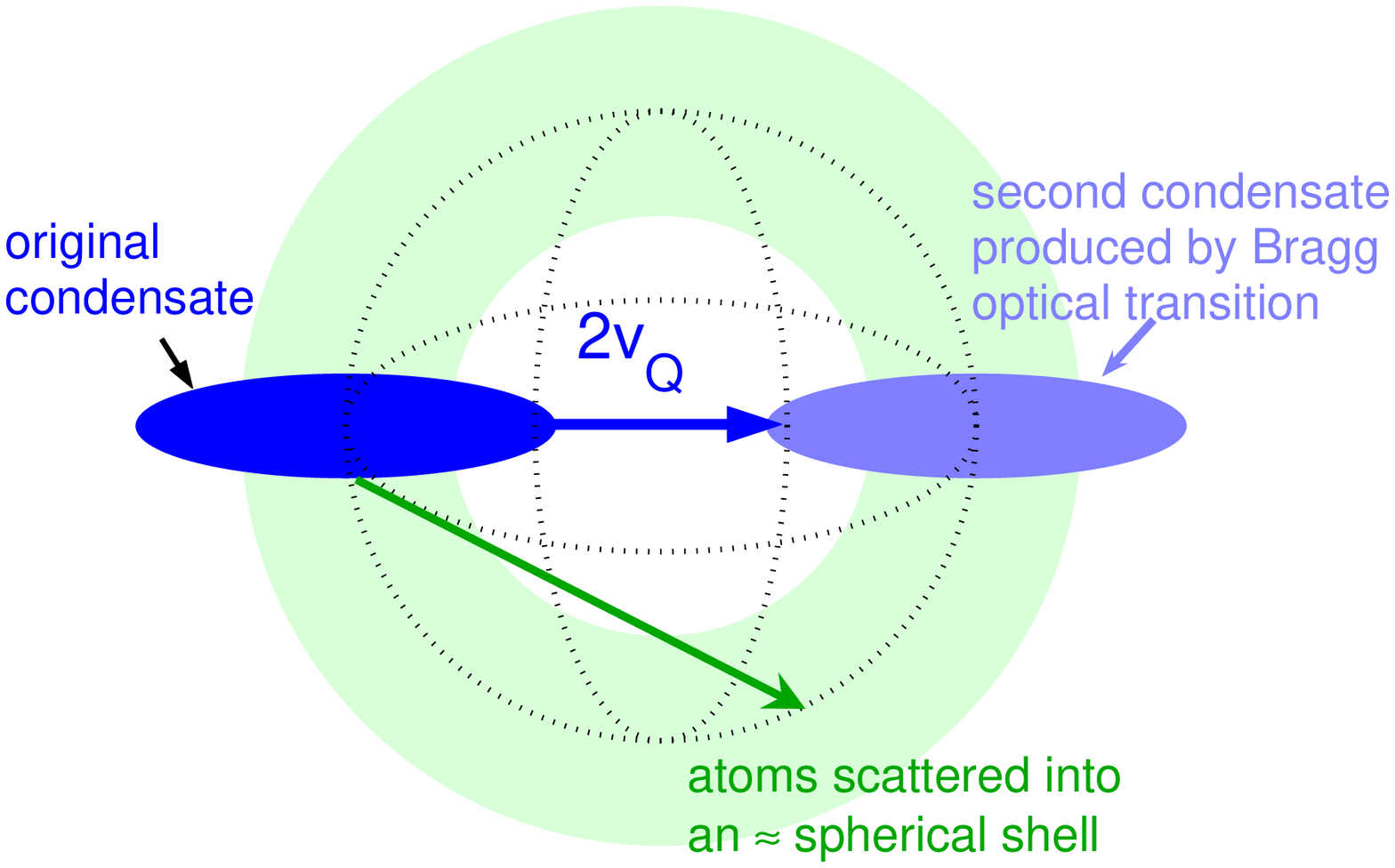}}\nosh{\raisebox{5cm}{\textbf{(a)}}}\hspace*{\columnwidth}\nosh{\mbox{}}\\
\nosh{\includegraphics*[width=\columnwidth]{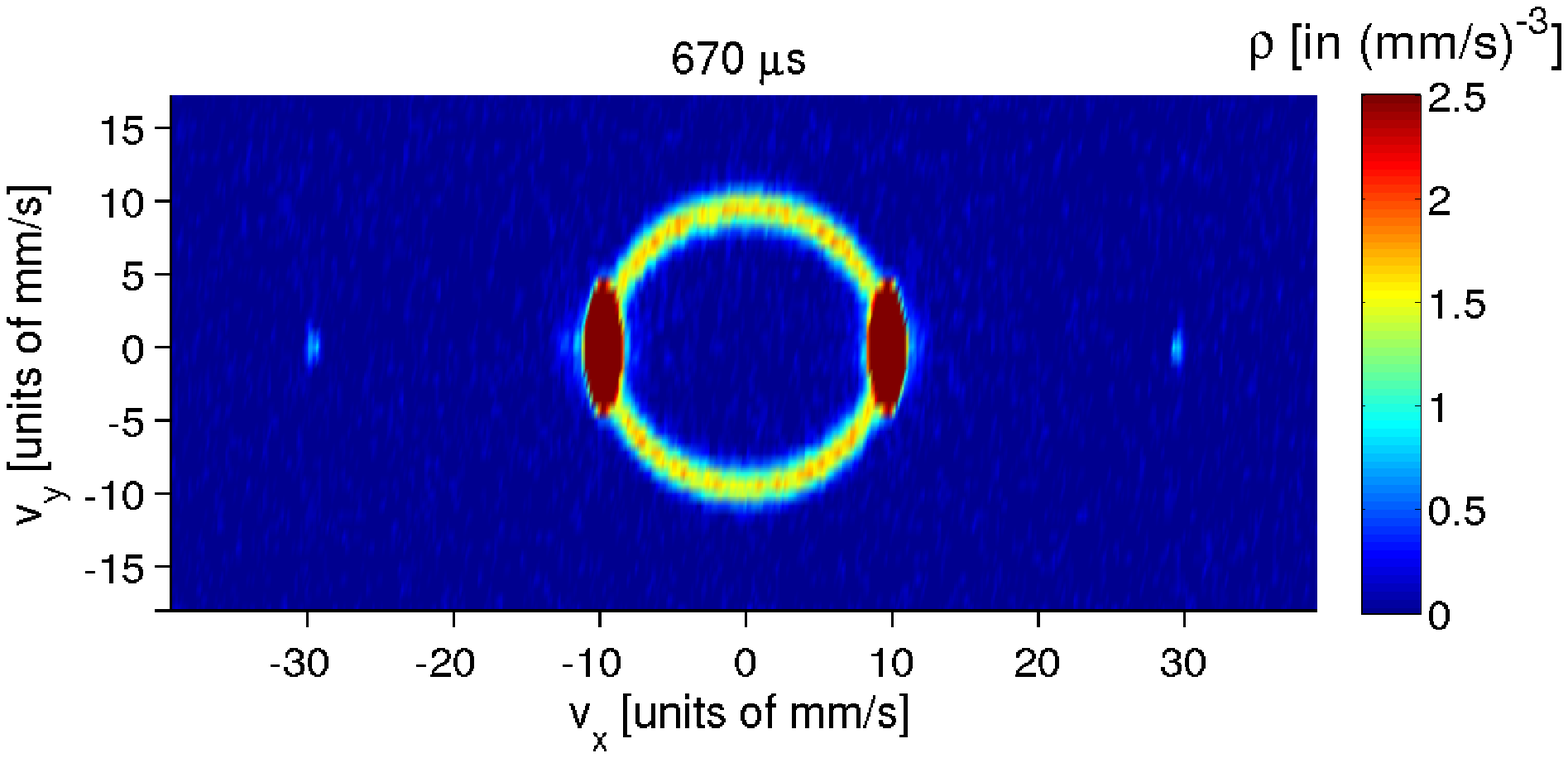}}\nosh{\raisebox{3.7cm}{\textbf{(b)}}}\hspace*{\columnwidth}\nosh{\mbox{}}

\noindent\small
The system simulated.
(a): Schematic of the BEC collision in real space in the lab frame. 
(b): Slice of the velocity distribution $\rho$ in the center-of-mass frame at $v_z=0$ and $t=670\mu$s calculated using the positive-P method.
This is about a third of the collision time, and the maximum time achievable with that method. 
The condensates are located around $v_x=\pm v_Q=\pm9.82$mm/s. 
The halo of scattered atoms is clearly seen, as are the coherent frequency doubling peaks at $\pm3v_Q\approx 30$mm/s. The collision is 
along the $x$ axis.
\end{minipage}
\vspace*{5cm}
\mbox{}

\clearpage
\newpage
\subsection{The relationship of the hybrid $\mc{H_A}$ to $s$-ordered operators}

First, a brief exposition of the standard formalism used in deriving phase-space quantum dynamics will be necessary. 
Writing the state of the system as a density matrix $\op{\rho}$, it can also be expressed as a distribution 
\begin{equation}\label{rhodef}
\op{\rho} = \int d\vec{v} P(\vec{v})\op{\Lambda}(\vec{v}).
\end{equation}
over a family of basis operators $\op{\Lambda}(\vec{v})$ parameterised by variables in the set $\vec{v}$. If the distribution
$P(\vec{v})$ is real and non-negative, this corresponds, in turn, to an ensemble of $\mc{S}$ sets of random variables $\vec{v}$ 
(``configurations'') chosen 
according to the distribution $P$, in the limit when $\mc{S}\to\infty$. In practice one 
computes a finite but large ensemble ($\mc{S}\gg1$) and knows properties of $\op{\rho}$ to 
within a statistical uncertainty that can be confidently estimated from the properties of the finite ensemble. 

The dynamics of the system is described by the master equation 
\begin{equation}\label{master}
i\hbar\pderiv{\op{\rho}}{t} = \left[\op{H}, \op{\rho}\right],
\end{equation}
while expectation values of observables are
\begin{equation}\label{obs}
\langle\op{O}\rangle = \tr\left[\op{O}\op{\rho}\right].
\end{equation}
These are most readily related to the computational ensemble of random variables through the use of the ``operator identities'', 
that are specific to each formulation.

For example, in the positive-P method one chooses $\op{\Lambda}$ to be an off-diagonal coherent-state operator. Letting 
$\bo{x}$ label discrete points in the computational lattice with $\Delta V$ volume per point, defining 
$$
\alpha_j(\bo{x}) = \psi_j(\bo{x})/\sqrt{\Delta V},
$$ 
one has
\begin{equation}
\op{\Lambda}_{PP}(\vec{v}) = \prod_{\bo{x}}\frac{|\alpha_1(\bo{x})\rangle_{\bo{x}}\langle\alpha_2(\bo{x})|_{\bo{x}}}
{\langle\alpha_2(\bo{x})|_{\bo{x}}|\alpha_1(\bo{x})\rangle_{\bo{x}}},
\end{equation}
where $\vec{v}=\left\{\alpha_1,\alpha_2\right\}$, 
$$
|\alpha\rangle_{\bo{x}} = e^{-|\alpha|^2/2}e^{\alpha\dagop{a}_{\bo{x}}}|0\rangle_{\bo{x}}
$$
is a coherent state on the $\bo{x}$ lattice point with the complex amplitude $\alpha$ and 
anihilation operator $\op{a}_{\bo{x}}=\op{\Psi}(\bo{x})\sqrt{\Delta V}$. Then, one finds 
(omitting ubiquitous local $\bo{x}$ dependence) the operator identities:
\begin{eqnarray*}
\op{\Psi}\op{\Lambda}_{PP}=\psi_1\op{\Lambda}_{PP} \quad&;& \quad
\dagop{\Psi}\op{\Lambda}_{PP}=\left[\psi_2^*+\pderiv{}{\psi_1}\right]\op{\Lambda}_{PP}\\
\op{\Lambda}_{PP}\dagop{\Psi}=\psi_2^*\op{\Lambda}_{PP} \quad&;&\quad 
\op{\Lambda}_{PP}\op{\Psi}=\left[\psi_1+\pderiv{}{\psi_2^*}\right]\op{\Lambda}_{PP},
\end{eqnarray*}
which are the source of the positive-P identities in the main text. Combined with (\ref{rhodef}) and (\ref{master})
these allow one to obtain a partial differential equation for $P(\vec{v},t)$ that is equivalent to the full quantum 
evolution of $\op{\rho}(t)$.
For the positive-P representation, this is a Fokker-Planck equation, and it corresponds exactly to the Langevin equations given in 
(5) of the main text 
Combining the identities with (\ref{obs}) and $\tr\left[\op{\Lambda}_{PP}\right]=1$ one finds 
$$
\langle\op{O}\rangle = \int P(\vec{v}) f_O(\vec{v}) d\vec{v}
$$
with a function $f_O$ that is obtained from $\op{O}$ via the operator identities, so that in the calculation it corresponds to 
an ensemble average  of $f_O$. For example, for $\op{O}=
\dagop{\Psi}(\bo{x})\op{\Psi}(\bo{x})$, the function is%
\footnote{$f_O= \psi_1^*(\bo{x})\psi_2(\bo{x})$ can also be obtained, but gives the same value of $\langle\op{O}\rangle$ 
in the $\mc{S}\to\infty$ limit.} 
 $f_O= \psi_2^*(\bo{x})\psi_1(\bo{x})$. 
The initial coherent 
state corresponds to $P = \prod_{\bo{x},j}\delta^{(3)}(\psi_j(\bo{x})-\phi_{GP}(\bo{x}))$.

It has been shown%
\footnote{K.~E.~Cahill and R.~J.~Glauber, Phys.~Rev.~\textbf{177}, 1857 (1969); \textsl{ibid.} \textbf{177}, 1882 (1969)} 
that the Glauber-Sudarshan P distribution described by a coherent state operator basis
$$
\op{\Lambda}_{GSP}(\psi) = \prod_{\bo{x}}|\alpha(\bo{x})\rangle_{\bo{x}}\langle\alpha(\bo{x})|_{\bo{x}}
$$
(similar to the positive-P but diagonal) 
can be described as the limit of a representation over $s$-ordered basis states
$$
\op{\Lambda}_{GSP} = \lim_{s\to1^{-}}\op{\Lambda}_s 
$$
where $s$ can take on continuous values from -1 to 1, and
\begin{equation}\label{opT}
\op{\Lambda}_s(\psi) = \prod_{\bo{x}}\frac{\op{D}(\alpha)_{\bo{x}}\op{T}(0,-s)_{\bo{x}}\op{D}^{-1}(\alpha)_{\bo{x}}}
{\tr\left[\op{D}(\alpha)_{\bo{x}}\op{T}(0,-s)_{\bo{x}}\op{D}^{-1}(\alpha)_{\bo{x}}\right]}.
\end{equation}
Here 
$$
\op{T}(0,-s)_{\bo{x}} = \frac{2}{1+s}\left(\frac{s-1}{1+s}\right)^{\dagop{a}_{\bo{x}}\op{a}_{\bo{x}}}
$$
is a kernel operator that becomes the vacuum $|0\rangle\langle0|$ in the limit of $s\to1^{-}$ and the local displacement operator is 
\begin{eqnarray*}
\op{D}(\alpha)_{\bo{x}} = e^{\alpha(\bo{x})\dagop{a}_{\bo{x}}-\alpha(\bo{x})^*\op{a}_{\bo{x}}}.
\end{eqnarray*}
so that coherent states are $|\alpha\rangle = \op{D}(\alpha)|0\rangle$. 
It was also shown there that the Wigner distribution corresponds to $s=0$, hence a variation of $s$ from 0 to 1 looks 
like a good candidate to create the $\mc{H_A}$ hybrid formulation between truncated Wigner and positive-P. The ``truncation'' 
refers to ad-hoc removal of third order\footnote{%
And higher order terms if necessary, although for the cold atom Hamiltonian considered in this letter, only partial derivatives up to third order are present 
in the Wigner representation.}
 partial derivatives of the Wigner distribution $P$ in its 
evolution equation to make it interpretable as Langevin stochastic equations of the samples. This removal is the reason 
why truncated Wigner treatments do not reproduce the full quantum dynamics.

First, though, one must take into account the off-diagonality that is responsible for the difference 
between the Glauber-Sudarshan P and positive-P: $\op{\Lambda}_{PP}\neq\op{\Lambda}_{GSP}$. 
Notably one of the bases%
\footnote{Though not the only one. Other ways of writing $\Lambda$ such as e.g. 
$\op{D}(\alpha_1)\op{T}(0,-1)\op{D}(\alpha_2^*)/\tr[\op{D}(\alpha_1)\op{T}(0,-1)\op{D}(\alpha_2^*)]$ 
can also reproduce the positive-P formulation but are not useful for generalisation to $s<1$, 
and do not reproduce the same itermediate operator identities.} that reproduces the positive-P is
\begin{eqnarray}
\op{\Lambda}_{PP}(\vec{v})&=&\prod_{\bo{x}}\frac{\op{d}(\vec{v})_{\bo{x}}\op{T}(0,-1)_{\bo{x}}\op{d}^{-1}(\vec{v})_{\bo{x}}}
{\tr\left[\op{d}(\vec{v})_{\bo{x}}\op{T}(0,-1)_{\bo{x}}\op{d}^{-1}(\vec{v})_{\bo{x}}\right]}\nonumber\\
&=& \prod_{\bo{x}}\op{d}(\vec{v})_{\bo{x}}\op{T}(0,-1)_{\bo{x}}\op{d}^{-1}(\vec{v})_{\bo{x}}\label{LambdaPP}
\end{eqnarray}
where the ``displacement-like'' operator
\begin{eqnarray*}
\op{d}(\vec{v})_{\bo{x}} = e^{\alpha_1(\bo{x})\dagop{a}_{\bo{x}}-\alpha_2(\bo{x})^*\op{a}_{\bo{x}}}.
\end{eqnarray*}
is obtained  by the replacement $\alpha\to\alpha_1,\alpha^*\to\alpha_2^*$ in $\op{D}(\alpha)$, and the second line follows because 
the trace in the denominator evaluates to one. The reason for this particular replacement is that for the positive-P 
distribution one requires $\op{\Lambda}$ to depend \textsl{analytically} on two separate complex variables, 
hence their complex conjugates must be removed. Here these analytic variables are $\alpha_1$ and $\alpha_2^*$. 

The extension of this $\op{\Lambda}$ onto a family of  $s$-ordered bases is 
\begin{eqnarray}
\op{\Lambda}_{s}^{\mc{A}}(\vec{v})&=&\prod_{\bo{x}}\frac{\op{d}(\vec{v})_{\bo{x}}\op{T}(0,-s)_{\bo{x}}\op{d}^{-1}(\vec{v})_{\bo{x}}}
{\tr\left[\op{d}(\vec{v})_{\bo{x}}\op{T}(0,-s)_{\bo{x}}\op{d}^{-1}(\vec{v})_{\bo{x}}\right]}\nonumber\\
&=& \prod_{\bo{x}}\op{d}(\vec{v})_{\bo{x}}\op{T}(0,-s)_{\bo{x}}\op{d}^{-1}(\vec{v})_{\bo{x}}.\label{HAbase}
\end{eqnarray}
This then interpolates towards the Wigner representation. Note that since the truncated Wigner 
evolution is deterministic, then if one takes the formally off-diagonal basis set with $s=0$ but 
imposes $\delta(\psi_1-\psi_2)$ in the initial conditions, it will remain exactly equivalent to the normal 
truncated Wigner formulation of (\ref{opT}) with $s=0$. 

One obtains the identities%
\footnote{For example, by comparison of expressions for LHS and RHS when $\op{T}(0,-s)$ is expanded in number states.}
\begin{eqnarray*}\label{HAop}
\op{\Psi}\op{\Lambda}^{\mc{A}}_s&=&\left[\psi_1-\frac{1-s}{2}\pderiv{}{\psi_2^*}\right]\op{\Lambda}^{\mc{A}}_s\\
\dagop{\Psi}\op{\Lambda}^{\mc{A}}_s&=&\left[\psi_2^*+\frac{1+s}{2}\pderiv{}{\psi_1}\right]\op{\Lambda}^{\mc{A}}_s \\
\op{\Lambda}^{\mc{A}}_s\dagop{\Psi}&=&\left[\psi_2^* -\frac{1-s}{2}\pderiv{}{\psi_1}\right]\op{\Lambda}^{\mc{A}}_s\\
\op{\Lambda}^{\mc{A}}_s\op{\Psi}&=&\left[\psi_1+\frac{1+s}{2}\pderiv{}{\psi_2^*}\right]\op{\Lambda}^{\mc{A}}_s
\end{eqnarray*}
which are exactly the same as was obtained by a naive blending of the operator 
identities in the main text provided we identify $\lambda=s$.

Regarding initial conditions, the diagonal $s$-ordered representation (\ref{opT}) for a coherent state $|\phi_{GP}\rangle$ was 
found by Cahill and Glauber to be Gaussian
\begin{equation}
P(\psi) = \prod_{\bo{x}} \frac{2}{1-s} \exp\left(-\frac{2|\psi(\bo{x})-\phi_{GP}(\bo{x})|^2}{\Delta V(1-s)}\right).
\end{equation}
When one additionally imposes $\psi_1=\psi_2=\psi$ as is done in the main text, this is equivalent to (\ref{HAbase}), 
justifying the initial conditions given in the main text that contain complex Gaussian noise of variance $(1-s)/2$.

\newpage
\subsection{Halo density calculations}

\begin{minipage}{\columnwidth}
\includegraphics[width=0.8\columnwidth]{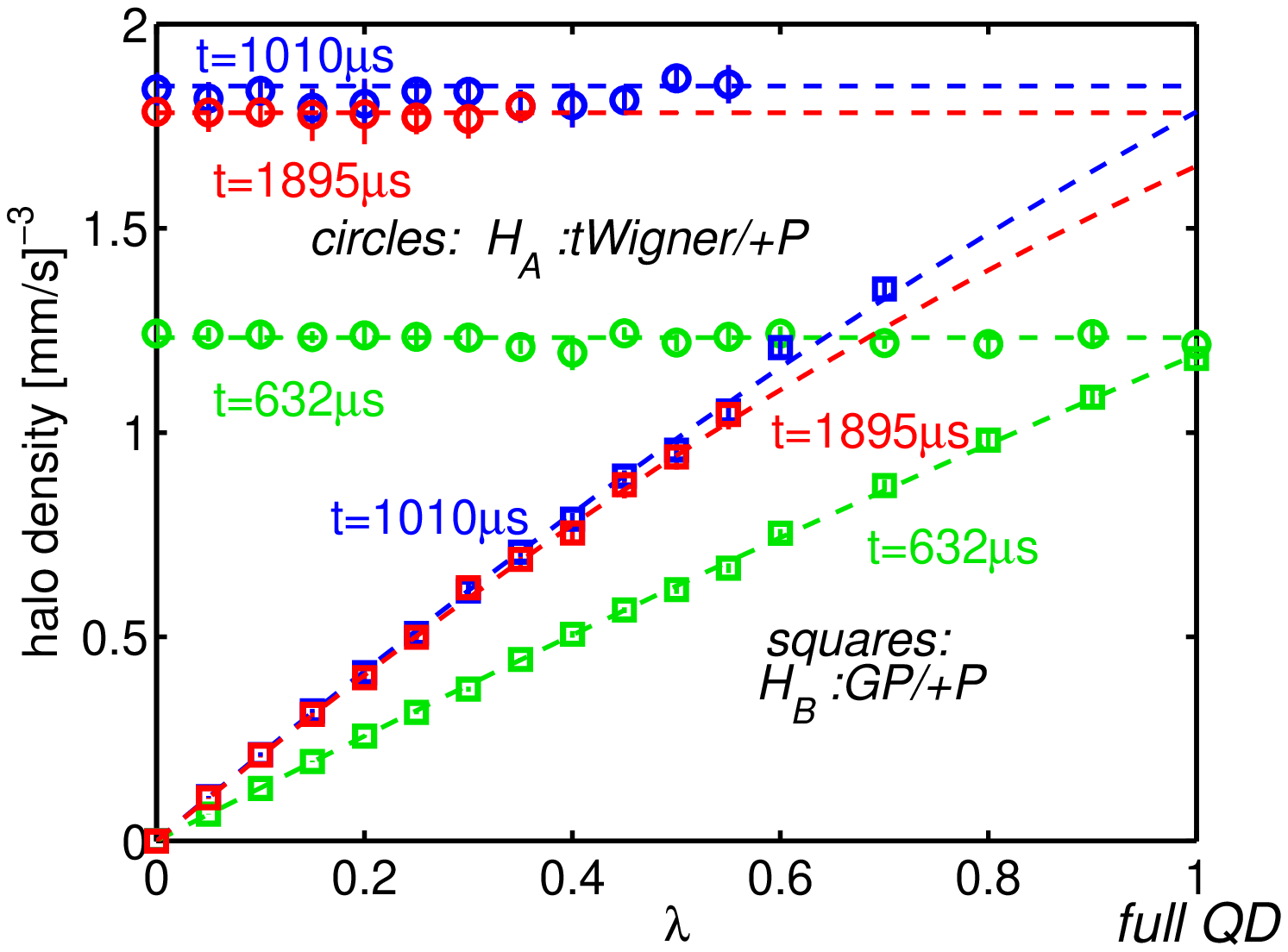}

\noindent\small
$\lambda$-dependent predictions of halo density (at $v_x=v_z=0$, $v_y=9.37$mm/s in velocity space) for several times (circles) with 
uncertainty shown as vertical bars at the same location. The corresponding fits (dashed) are quadratic for the $\mc{H_B}$ hybrid, and 
constant-value for $\mc{H_A}$. Fitting is via minimisation of 
rms deviation in units of $1\sigma$ data uncertainty. Linear or quadratic fits to the $\mc{H_A}$ hybrid data are not more statistically 
significant than the constant-value fit, and hence would be poorly conditioned. 
\end{minipage}

\begin{minipage}{\columnwidth}
\includegraphics[width=0.8\columnwidth]{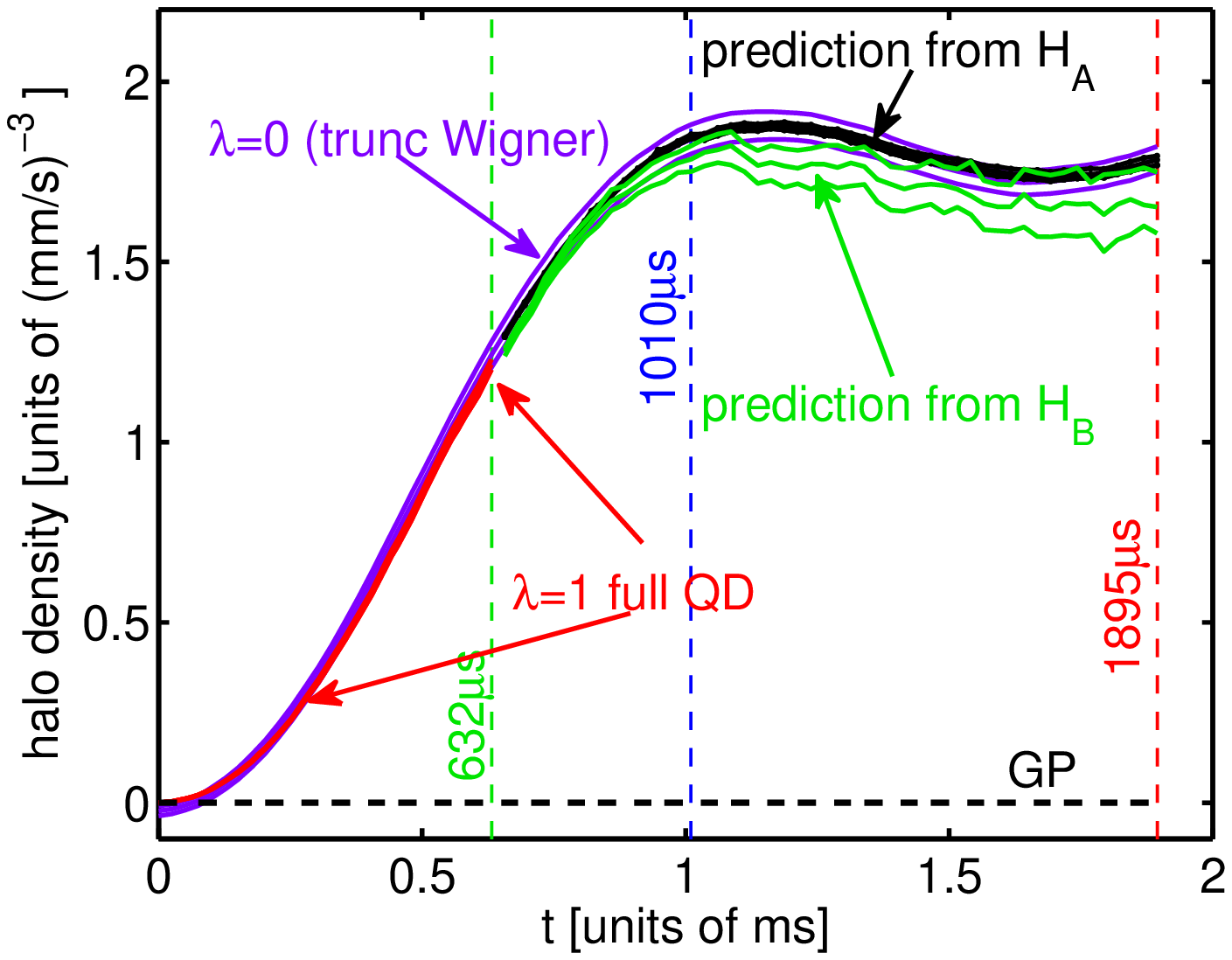}

\noindent\small
Predictions of halo density (at $v_x=v_z=0$, $v_y=9.37$mm/s in velocity space) from hybrids $\mc{H_A}$ and $\mc{H_B}$ compared with short-time full quantum dynamics 
and approximate methods. Triple lines, where visible, are $1\sigma$ uncertainty. Prediction data 
based on $\approx 10-20$ values of $\lambda$, each with $\approx300-1000$ trajectories, and quadratic / constant-value fitting for 
$\mc{H_A}$ / $\mc{H_B}$ hybrids, respectively. Note the agreement with truncated Wigner to within statistical uncertainty.
Times detailed in the previous figure (above) are highlighted.
\end{minipage}

\vfill
\mbox{}

\subsection{Extrapolation from partial $\lambda$ segment}

\begin{minipage}{\columnwidth}
\includegraphics[width=0.8\columnwidth]{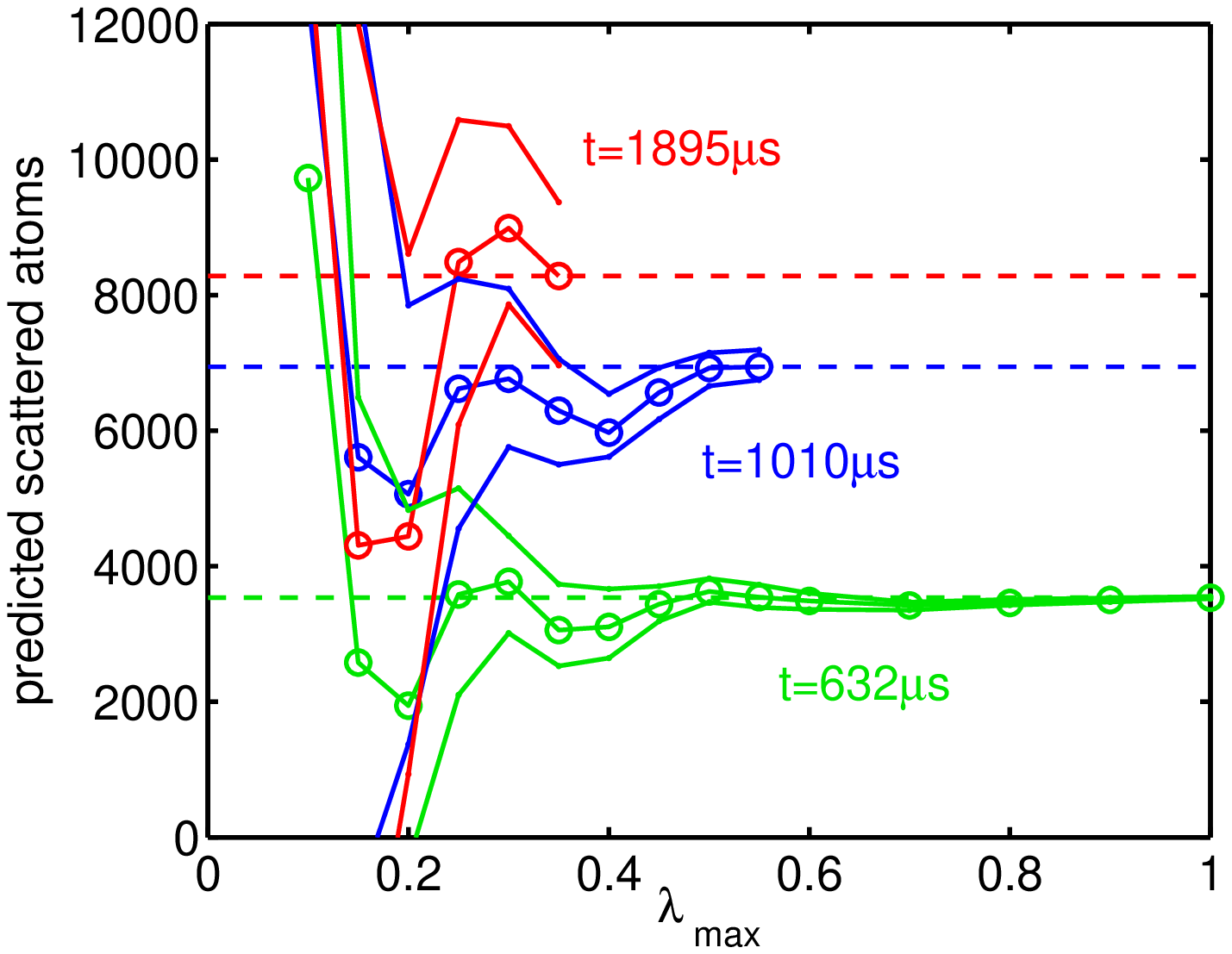}

\noindent\small
Predictions of the number of scattered atoms at several times, as a function of the $\lambda$ segment $\lambda\in[0, \lambda_{\rm max}]$ 
used for extrapolation from a quadratic fit to $\mc{H_A}$ results.  Triple lines, where visible, are $1\sigma$ uncertainty. Dashed lines indicate
the final predictions using all the available $\lambda$ values. Data used was from the same simulations as in 
Fig.~2
of the main text. There is no statistically significant trend with $\lambda_{\rm max}$ visible, 
suggesting that the fitting function that is a quadratic polynomial in $\lambda$ is appropriate within statistical precision.
\end{minipage}

\clearpage

\end{document}